\documentclass[aps,superscriptaddress,twocolumn,twoside,floatfix,prl,nofootinbib,a4paper]{revtex4-1}
\usepackage{graphicx,amsmath,amssymb,color}
\usepackage[normalem]{ulem}
\usepackage{amsfonts}
\usepackage[ocgcolorlinks,colorlinks=true,linkcolor=blue,citecolor=red]{hyperref}
\usepackage{latexsym}
\usepackage{amsfonts}
\usepackage{mathrsfs}
\usepackage{natbib}
\usepackage{color,verbatim}
\DeclareMathAlphabet{\mathrsfs}{U}{rsfs}{m}{n}
\DeclareMathAlphabet{\mathpzc}{OT1}{pzc}{m}{it}
\DeclareMathAlphabet{\matheus}{U}{eus}{m}{n}
\DeclareMathAlphabet{\mathbbold}{U}{bbold}{m}{n}

\newcommand{\ba}{\begin{eqnarray}}
\newcommand{\be}{\begin{equation}}
\newcommand{\ee}{\end{equation}}

\newcommand{\ea}{\end{eqnarray}}
\newcommand{\ban}{\begin{eqnarray*}}
\newcommand{\ean}{\end{eqnarray*}}

\newcommand{\ket}[1]{|#1\rangle}

\begin{document}

\title{Quantum Random Access Codes using Single d-level Systems}
\author{Armin Tavakoli}
\affiliation{Department of Physics, Stockholm University, S-10691 Stockholm, Sweden}

\author{Alley Hameedi}
\affiliation{Department of Physics, Stockholm University, S-10691 Stockholm, Sweden}

\author{Breno Marques}
\affiliation{Department of Physics, Stockholm University, S-10691 Stockholm, Sweden}

\author{Mohamed Bourennane}
\affiliation{Department of Physics, Stockholm University, S-10691 Stockholm, Sweden}

\begin{abstract}
Random access codes (RACs) are used by a party to despite limited communication access an arbitrary subset of information held by another party. Quantum resources are known to enable RACs that break classical limitations. Here, we study quantum and classical RACs with high-level communication. We derive average performances of classical RACs and present families of high-level quantum RACs. Our results show that high-level quantum systems can significantly increase the advantage of quantum RACs over the classical counterparts. We demonstrate our findings in an experimental realization of a quantum RAC with four-level communication.

\end{abstract}

\maketitle

\textit{Introduction.---} Quantum information arises from the realization that information can be encoded in physical systems subject to the laws of quantum mechanics, and thus allowing for the use of non-classical phenomena such as superpositions and entanglement as quantum resources for information processing. Today, it is well-known that quantum resources can enhance information processing beyond classical limitations in cryptography \cite{BB84, E91, CBKG02, GRTZ02}, communication complexity \cite{CG97, G01, BCMW10} and numerous other information or computation problems \cite{FGM01, TC15, G96, S94}.

A broad category of communication tasks that are useful for a wide variety of applications is random access codes (RACs). In a RAC, one party Alice holds a set of information and another party Bob aims to access an arbitrary subset of the information held by Alice. However, the communication between the parties is restricted. We can denote a RAC as $n\overset{p}{\rightarrow}1$, meaning that Alice has $n$ bits of information while Bob is interested in one of Alice's bits. The task is for Alice to encode her $n$ bits into only one bit, which she sends to Bob, such that Bob can find the value of any of her bits with a high probability $p$.

In the analog quantum random access code (QRAC), we allow Alice to encode her $n$ bits into a single two-level quantum system that she sends to Bob. Depending on which of Alice's bits Bob wishes to find, he performs a suitable measurement on the system and with high probability recovers the true value of the particular bit. One should distinguish between the following two scenarios: Bob recovers the bit of is interest with (i) a worst case probability $p$, and (ii) an average probability $p$.

QRACs were initially introduced in studies of quantum finite automata \cite{ANTV99, ANTV02, N99}. Nevertheless, QRACs were shown useful for various other applications including locally decodable codes \cite{K04}, network coding \cite{3}, reduction of communication complexity \cite{G02}, semi-device independent random number expansion \cite{LYW11} and semi-device independent key distribution \cite{PB11}. Additionally, QRACs are interesting for foundational studies of quantum and no-signaling resources \cite{IC09, GH14}.  Furthermore, random access coding can be realized by means of entanglement and classical communication \cite{PZ10} and such entanglement-assisted RACs have been applied to enhance strategies in games \cite{MT14}.

In \cite{ANTV99}, an example of a $2\overset{0.854}{\rightarrow}1$ QRAC was presented which provides an advantage over the corresponding classical RAC. Classically, one can achieve an average success probability $p^C=3/4$ which can be obtained as follows. Let Alice send Bob the value of her first bit every time. Whenever Bob is interested in this bit he recovers it with certainty and whenever he is interested in Alice's second bit he is forced to guess, succeeding in half the cases,  leading to $p^C=3/4$. However, in the QRAC, Alice encodes her two bits $x_0,x_1\in\{0,1\}$ into a two-level system $
|\psi_{x_0x_1}\rangle=\frac{1}{\sqrt{2}}\left(|0\rangle+\frac{1}{\sqrt{2}}\left((-1)^{x_0}+i(-1)^{x_1}\right)|1\rangle\right)$. If Bob is interested in $x_0$ he performs a measurement in the basis $\{\frac{1}{\sqrt{2}} \left(1,1\right),\frac{1}{\sqrt{2}} \left(1,-1\right)\}$ and if he is interested in $x_1$ he measures in the basis $\{\frac{1}{\sqrt{2}} \left(1,i\right),\frac{1}{\sqrt{2}} \left(1,-i\right)\}$. Since the encoding states are symmetrically distributed on the equator of the Bloch sphere such that they form the vertices of a square, the probabilities of all outcomes are the same, namely $p^Q=\frac{1}{2}\left(1+1/\sqrt{2}\right)\approx 0.854$. Thus, the QRAC outperforms the RAC, and we quantify the advantage by the ratio $p^Q/p^C\approx 1.138$.

As mentioned in \cite{ANTV99}, the above example was extended by Chuang to a $3\overset{0.789}{\rightarrow}1$ QRAC using eight encoding states symmetrically distributed over the surface of the Bloch sphere forming the vertices of a cube, and adding the third measurement basis $\{\left(1,0\right),\left(0,1\right)\}$. Due to the symmetry, the success probability is always $p^Q=1/2(1+1/\sqrt{3})\approx 0.789$ whereas classically, one can still achieve $p^C=3/4$ with a somewhat more sophisticated strategy than in the $2\rightarrow 1$ case. Hence, the advantage of this QRAC over the RAC is $p^Q/p^C\approx 1.052$.

However, RACs (QRACs) can also be considered when Alice and Bob do not use bits (qubits) but a classical (quantum) $d$-level system. To the authors' knowledge, little attention has been directed to such generalizations but initial steps have been taken in \cite{G02, CGS08}. In this letter, we study RACs and QRACs with high-level communication, and we infer the notation $n^{(d)}\overset{p_{n,d}}{\rightarrow} 1$ as an abbreviation of the scenario where Alice has $n$ $d$-levels i.e. a string  $x=x_0...x_{n-1}$ where $x_i\in\{0,...,d-1\}$, which she encodes into a classical (quantum) $d$-level, which is sent to Bob who should be able to recover the $x_j$ of his interest with an average (or worst case) probability of $p_{n,d}$.

\textit{Classical RACs.---} Let us begin by considering classical RACs on the form  $n^{(d)}\overset{p_{n,d}}{\rightarrow}1$. Any deterministic classical strategy for the RAC is composed of an encoding function used by Alice to encode her string into a $d$-level, and a decoding function used by Bob to decode the obtained $d$-level with respect to $x_j$. Proving the optimality of a classical strategy is difficult for the general $n^{(d)}\rightarrow 1$ setting. However, intuition strongly suggests that an optimal strategy is for Alice to use majority encoding i.e. Alice counts the number of times each of the values $\{0,...,d-1\}$ appears in her string $x$ and outputs the $d$-level that occurs most frequently. In case of a tie, she can output either $d$-level. Bob does identity decoding and thus outputs whatever he receives from Alice. This strategy was proven optimal for $d=2$ in \cite{ALMO}.

Now, we find the average classical success probability $p_{n,d}^C$ for arbitrary $n$ and $d$. Let all positive integer partitions of the number $n$ be stored into the set $X$ and let $X_j$ denote the $j$'th partition. The total number of partitions is $|X|$. In total, there are $d^n$ strings that Alice could be holding, and for each string $x$ we make a frequency list i.e. a table where we count the number of times any same $d$-level occurs in $x$. Thus, for any given $x$ there exists a unique $j$ such that $x$ associates to the partition $X_j$: $x\sim X_j$. Consider the number of strings $N_{X_j}$ such that $x\sim X_j$. We compute $N_{X_j}$ combinatorially by
\begin{equation}\label{NXj}
N_{X_j}=\frac{n!}{k_1!...k_{|X_j|}!}\left(\prod_{l\in \overline{X}_j} C_{X_j}(l)!\right)^{-1}\prod_{m=1}^{|X_j|}(d-m+1)
\end{equation}
where $k_1,...,k_{|X_j|}$ are the elements of $X_j$, $\overline{X}_j$ is the set of distinct elements occurring in $X_j$ and $C_{X_j}(l)$ is the number of times $l$ appears in $X_j$. Given an "empty" string $x$ with $n$ positions we need to assign one of $d$ possible values to the first $k_1$ positions in $x$. Similarly, we can assign any of the remaining $d-1$ values to the next $k_2$ positions in $x$. This process is continued until we have assigned one of the remaining $d-|X_j|+1$ values to the final $k_{|X_j|}$ positions in $x$. The number of such strings is given by the third factor in \eqref{NXj}. However, if we arbitrarily permute $x$ into $x'$, it still holds that $x'\sim X_j$, and therefore we multiply by the first factor in \eqref{NXj} which is the number of permutations of $x$. Nevertheless, if it happens that $k_i$ occurs more than once in $X_j$, some strings will be counted $C_{X_j}(k_i)!$ times. By multiplication with the second factor in \eqref{NXj}, each string is counted only once and thus \eqref{NXj} is the number of strings $x\sim X_j$.

Using majority encoding, Alice will compute the maximum frequency $\max \{X_j\}$ and output the associated $d$-level to Bob who's identity decoding has a probability of $\max \{X_j\}/n$ to return the correct value of the $d$-level of his interest. Therefore, the average success probability is
\begin{equation}\label{RAC}
p_{n,d}^C=\frac{1}{nd^n}\sum_{j=1}^{|X|}\max\{X_j\}N_{X_j}
\end{equation}
Evaluating this expression for the general $(n,d)$-setting is cumbersome. Therefore we have computed two special cases by fixing $n=2,3$
\begin{equation}\label{C2}
p_{2,d}^C=\frac{1}{2}\left(1+\frac{1}{d}\right)\hspace{7 mm}
p^C_{3,d}=\frac{1}{3}\left(1+\frac{3}{d}-\frac{1}{d^2}\right)
\end{equation}
Observe that in the limit of $d\rightarrow \infty$ the classical success probability converges to $1/n$ which corresponds to the trivial strategy where Bob simply guesses which of the $n$ $d$-levels he was given.

\textit{Quantum advantage of QRACs in \cite{CGS08}.---}
Let us now shortly apply our formula \eqref{RAC} to investigate the advantage of some known high-level QRACs. In \cite{CGS08}, a set of QRACs on the form $(d+1)^{(d)}\rightarrow 1$ for $d=2,3,4,5,7,8$ were constructed from states maximizing the discrete Wigner function. The authors compute the average success probability of their QRACs but are unable to compare these with the success probability of the corresponding RAC. Having found the success probability of the RACs, we can implement our results with the QRACs of \cite{CGS08} to quantify the quantum advantage, see table \ref{tab:1}.
\begin{table}[h]
\centering
\begin{tabular}{| c| c| c| c|}
\hline
 $ d $ & $p^Q$ in \cite{CGS08}& $p^C$ & $p^Q/p^C$ \\ [1ex] 
\hline 
2 & $ 0.789$ & $0.75$ & $1.052$ \\ 
3 & $ 0.637$ & $ 0.593$ & $ 1.075$ \\
4 & $ 0.5424$ & $0.4961$ & $ 1.0933$ \\
5 & $ 0.4700$ & $0.4291$ & $1.0953$ \\
7 & $ 0.3720$ & $0.3420$ & $1.0876$ \\
8 & $ 0.3372$ & $0.3118$ & $1.0815$ \\[1ex] 
\hline 
\end{tabular}
\caption{Comparison of quantum and classical success probabilities for QRACs in \cite{CGS08}.}
\label{tab:1}
\end{table}
In the case of $d=2$ the result reproduces the known $3^{(2)}\rightarrow 1$ QRAC of Chuang. However, we see that the QRACs with $n=d+1>2$ can provide an advantage larger than the previously known case of $d=2$.

\textit{High-level QRACs with $n=2$.---} We will now construct a family of QRACs on the form $2^{(d)}\rightarrow 1$ i.e. Alice has $x_0,x_1\in\{0,...,d-1\}$ which she encodes into a quantum $d$-level system sent to Bob. For this purpose we will use two mutually unbiased bases (MUBs) \cite{I81, WF89}, namely the computational basis $\{|l\rangle\}_l$ and the fourier basis $|e_l\rangle=\frac{1}{\sqrt{d}}\sum_{k=0}^{d-1}\omega^{kl}|k\rangle$ where $\omega=e^{\frac{2\pi i}{d}}$. Let us assume that Alice encodes the string $x_0x_1=00$ into the state $|\psi_{00}\rangle=\frac{|0\rangle+|e_0\rangle}{N_{2,d}}$ where $N_{2,d}=\sqrt{2+\frac{2}{\sqrt{d}}}$ is the $d$-dependent normalization. We now use the unitary operators $X=\sum_{k=0}^{d-1}|k+1\rangle\langle k|$ and $Z=\sum_{k=0}^{d-1}\omega^{k}|k\rangle \langle k|$
to define the general encoding state $|\psi_{x_0x_1}\rangle$ as
\begin{equation}\label{2to1enc}
|\psi_{x_0x_1}\rangle=X^{x_0}Z^{x_1} |\psi_{00}\rangle
\end{equation}
Bob will perform a measurement in the basis $\{|l\rangle\}_l$ if he is interested in $x_0$ and in the basis $\{|e_l\rangle\}_l$ if he is interested in $x_1$. We proceed by deriving the success probability $p_{2,d}^Q$ of this family of QRACs.

Firstly, we expand the encoding state \eqref{2to1enc} onto the form $|\psi_{x_0x_1}\rangle=\frac{1}{N_{2,d}}\left(|x_0\rangle+\frac{1}{\sqrt{d}}\sum_{j=0}^{d-1}\omega^{jx_1}|j+x_0\rangle \right)$. Then, if Bob  performs a computational basis measurement and his outcome is denoted by $l$, the probability distribution $P_0$ of Bob's outcomes is
\begin{equation}\label{2to1comp}
P_{0}(l)\equiv|\langle l|\psi_{x_0x_1}\rangle|^2=\frac{1}{N_{2,d}^2}\left|\delta_{l,x_0}+\frac{\omega^{x_1(l-x_0)}}{\sqrt{d}}\right|^2
\end{equation}
On the other hand, assume that Bob is interested in $x_1$, then he will measure in the fourier basis giving rise to a probability distribution $P_1$,
\begin{equation}\label{2to1fourier}
P_{1}(l)\equiv|\langle e_l|\psi_{x_0x_1}\rangle|^2=\frac{1}{N_{2,d}^2}\left|\frac{\omega^{-lx_0}}{\sqrt{d}}+ \omega^{-x_0x_1}\delta_{x_1,l} \right|^2
\end{equation}
Bob's probability of success amounts to his measurement outcome indeed being the correct value of the $x_j$ he is interested in. This corresponds to the outcome $l=x_0$ in \eqref{2to1comp} and the outcome $l=x_1$ in \eqref{2to1fourier}. Evaluating both these probabilities expressions will yield
\begin{equation}\label{QRAC}
p_{2,d}^Q=\frac{1}{2}\left(1+\frac{1}{\sqrt{d}}\right)
\end{equation}
For any $d$ the QRAC is non-trivial since $p_{2,d}^Q>1/2$. In the case of $d=2$ the well known result of $p_{2,2}^Q\approx 0.854$ is reproduced. Furthermore, since the success probability of Bob only depends on $d$, the proposed family of QRACs perform equally well both on average and in the worst case setting i.e. the probability to recover $x_0$ is always the same as the probability to recover $x_1$, independent of the encoding or the measurement.

\begin{figure}
\centering
\includegraphics[width=1\columnwidth]{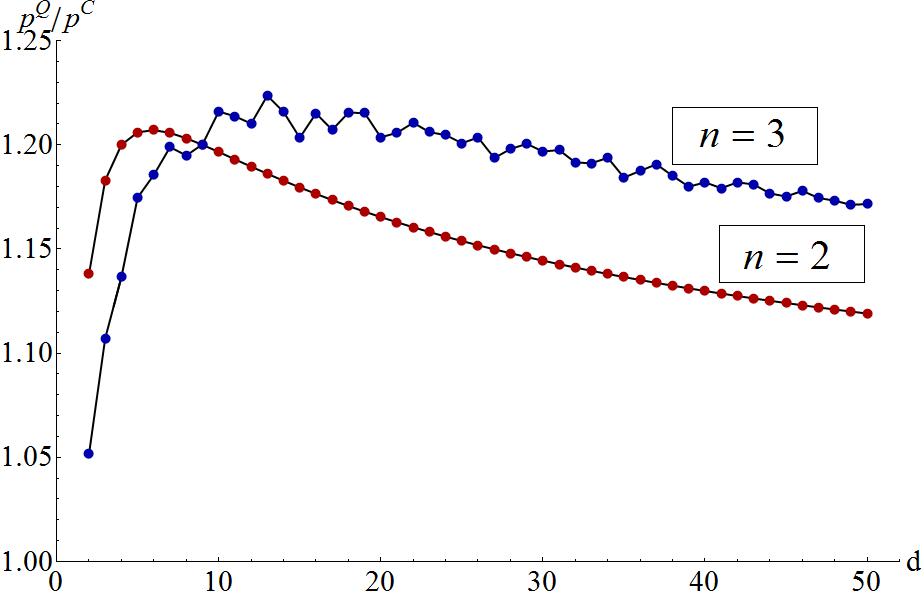}
\caption{The ratio between the quantum and classical success probabilities as a function of $d$ for the $2^{(d)}\rightarrow 1$ (red) and $3^{(d)}\rightarrow 1$ (blue) QRACs.}\label{fig:1}
\end{figure}

The quantum success probability \eqref{QRAC} can be compared to the classical success probability \eqref{C2}. Evidently, it holds that $p_{2,d}^Q>p_{2,d}^C$ for all $d$. We quantify the advantage of the QRAC over the RAC by the ratio $p^Q/p^C$, and in figure \eqref{fig:1}, we represent the quantum advantage by the red dots. The optimal quantum advantage is found at $d=6$ at which $p_{2,6}^Q/p_{2,6}^C=\frac{1}{7}\left(6+\sqrt{6}\right)\approx 1.207$. This is constitutes a significant improvement over the $d=2$ case where the quantum advantage is only about $1.138$. In terms of absolute numbers, the difference between the quantum and the classical success probability is maximized for $d=4$ where $p^Q_{2,4}-p^C_{2,4}=1/8$.

\textit{Experimental realization.--} Now, we will now present an experimental realization of the $2^{(4)}\rightarrow 1$ QRAC.
In our experiment, the physical systems are defined by single photons in polarization and path setup. The information is encoded in  four basis states: $|1\rangle \equiv |H,a\rangle$, $|2\rangle \equiv |V,a\rangle$, $|3\rangle \equiv |H,b\rangle$ and $|0\rangle \equiv |V,b\rangle$, where   ($H$) and  ($V$) are horizontal and vertical polarization photonic modes respectively, and ($a$ and $b$) are two spatial photonic modes of single photons. Any ququart state can be written as $\alpha |H,a\rangle + \beta |V,a\rangle + \gamma |H,b\rangle + \delta |V,b\rangle$.

\begin{figure}[t]
\centerline{\includegraphics[width=1\columnwidth]{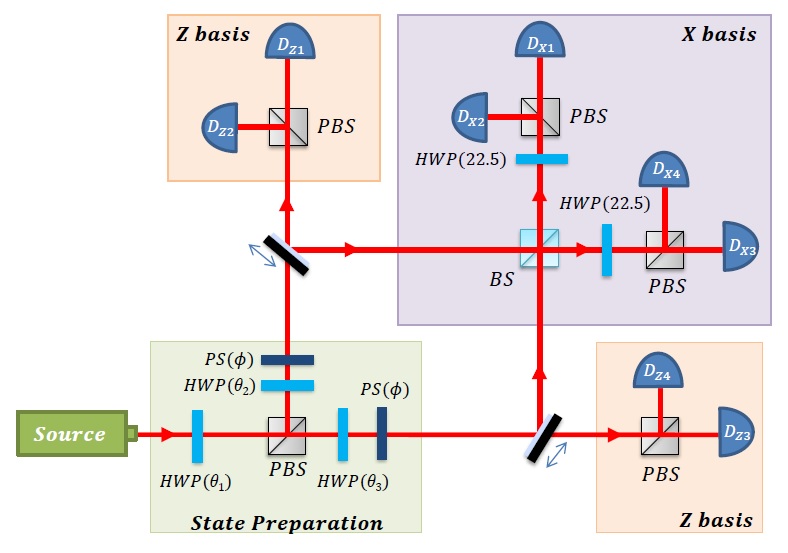}}
\caption{Experimental set-up for the realization of four-level encoding QRAC. Alice's states are encoded in single photon horizontal and vertical polarizations, and in two spatial modes and prepared by suitable orientation of three half-wave plates (HWP) at angles $\theta_i$ (with $i=1,2,3$), PBS and phase shift PS ($\phi$). When Bob measures in the computational basis, the measurement consists, in each of two spacial modes (a) and (b), of PBS and two single photon detectors $D_{Zi}$ ($i=1,...4$). Bob's measurement in the Fourier basis consists of an interference at BS and at each of the outputs of the BS, of HWP oriented at 22.5, PBS, and two single photon detectors $D_{Xi}$ ($i=1,...4$) .}
\label{Fig2}
\end{figure}

The single photons are generated from a heralded single photon source using a 390~nm pulsed laser pumping a 2~mm BBO crystal and creating twin photons through a spontaneous parametric down-conversion (SPDC) process. The idler photon is used as trigger and detected by a single photon detector $D_T$. To exactly define the spatial and spectral properties of the signal photon, the emitted photon modes are coupled into a single mode fiber (SMF) and passed through a 3~nm narrowband interference filter (F). We have characterized our heralded single photon source and we have found that the ratio between the coincidence photon counting due to single- and multi-photon pair emissions is below  $0.1\%$. To prepare the initial state of the photon in $\ket{H}$, the signal photon is passing through a polarizer oriented to horizontal polarization direction. Alice's photonic states were prepared by three suitably oriented half-wave plates HWP($\theta_1$), HWP($\theta_2$) and HWP($\theta_3$), polarization beam splitter (PBS),  and  a setting of a phase shift PS($\phi$) (see figure \ref{Fig2}), $cos(2\theta_1)cos(2\theta_2) |H,a\rangle + cos(2\theta_1)sin(2\theta_2) |V,a\rangle + e^{i\phi}\left[sin(2\theta_1)sin(2\theta_3) |H,b\rangle - sin(2\theta_1)cos(2\theta_3) |V,b\rangle\right]$. Thus, by adjusting the HWP orientation angles $\theta_i$, Alice could produce any of the 16 required states $|\psi_{x_0x_1}\rangle$ with $x_0,x_1\in \{0,1,2,3\}$, see Table \ref{table2} for all settings.

Bob chooses between two measurement settings, either in the basis of eigenstates of $Z$ (computational basis) or $X$ (Fourier basis). For the computational basis, the measurement consists in each of two spatial modes (a) and (b), of PBS and two single photon detectors $D_{Zi}$ ($i=1,...4$). The choice to measure in a particular basis is implemented by moving the mirrors ($M$) in and out with help of pico motors translation stages. The Fourier basis measurement consists of an interference at BS and at each of the outputs of the BS, of HWP oriented at $22.5$, PBS, and two single photon detectors ($D_{Xi}$ ($i=1,...4$).

Our single-photon detectors ($D_i$, $i=1,2,3,4$) for the two measurements were Silicon avalanche photodiodes with detection efficiency $\eta_d = 0.55$, the dead  time $50$ ns, and dark counts rate $R_d \simeq 400$ Hz. All coincidence counts between the signal and idler photons are registered using an Multi-channel coincidence logic with a time window of $1.7$~ns. The number of detected photons was approximately $2500$ per second and the total time used for each experimental settings was $10$~s.

The probabilities needed for the QRAC are obtained from the number of detections in the single photon detectors.
All experimental results are presented in Table~\ref{table2}. The results are in very good agreement with the predictions of quantum mechanics, namely for an ideal experiment $P^Q = 0.75$. Our results yield an average quantum success probability $0.754 \pm 0.038$ which violates the classical bound $P^C = 0.625$. The errors come from Poissonian counting statistics and systematic errors. The main sources of systematic errors are the slight intrinsic imperfections of the PBSs and HWPs.

\begin{table}[t]
\begin{tabular}{|c|cccc|c|c|} \hline\cline{1-7}

$ \psi_{x_0x_1} $&$\theta_1$ & $\theta_{2}$ & $\theta_3$ & $\phi$& $P^Z_{exp}$ & $P^X_{exp}$ \\
\hline\cline{1-7}

\: $\psi_{00}$ & 12.05 & 22.5  & -9.22  & 0  & $0.747\pm 0.036$ & $0.752\pm 0.038$\\
\: $\psi_{01}$ & 12.05 & 22.5  & -9.22  & $\pi$  & $0.748\pm 0.036 $ & $0.774\pm 0.038$    \\
\: $\psi_{02}$ & 12.05 & -22.5 & 9.22   & 0  &  $0.749\pm 0.036$ & $0.752\pm 0.039$ \\
\: $\psi_{03}$ & 12.05 & -22.5 & 9.22   & $\pi$  &  $0.753\pm 0.036 $ & $0.713\pm 0.039$ \\
\: $\psi_{10}$ & 12.05 & 22.5  & -35.78 & 0 &  $0.766 \pm 0.036 $ & $0.763\pm 0.038$ \\
\: $\psi_{11}$ & 12.05 & 22.5  & -35.78 & $\pi$ & $0.767\pm 0.036 $ & $0.750\pm 0.038$  \\
\: $\psi_{12}$ & 12.05 & -22.5 & 35.78  & 0  &  $0.764\pm 0.036 $ & $0.751\pm 0.039$ \\
\: $\psi_{13}$ & 12.05 & -22.5 & 35.78  & $\pi$ &  $0.764\pm 0.036 $ & $0.748\pm 0.039$  \\
\: $\psi_{20}$ & -22.5 & 32.95 & 9.22   & 0  & $0.755\pm 0.036 $ & $0.764\pm 0.038$ \\
\: $\psi_{21}$ & -22.5 & 32.95 & 9.22   & $\pi$ & $0.761\pm 0.036 $ & $0.754\pm 0.038$\\
\: $\psi_{22}$ & 22.5  & 32.95 & -9.22  &  0  & $0.786\pm 0.036 $ & $0.775\pm 0.039$\\
\: $\psi_{23}$ & 22.5  & 32.95 & -9.22  &  $\pi$ & $0.786\pm 0.036 $ & $0.728\pm 0.039$\\
\: $\psi_{30}$ & -22.5 & 32.95 & 35.78  &  0 & $0.742\pm 0.037 $ & $0.772\pm 0.038$\\
\: $\psi_{31}$ & -22-5 & 32.95 & 35.78  &  $\pi$ & $0.743\pm 0.037 $ & $0.766\pm 0.038$\\
\: $\psi_{32}$ & 22.5  & 32.95 & -35.78 & 0 & $0.732\pm 0.037 $& $0.749\pm 0.039$ \\
\: $\psi_{33}$ & 22.5  & 32.95 & -35.78 & $\pi$ & $0.733\pm 0.037 $ & $0.703\pm 0.039$\\

\hline\cline{1-7}
\end{tabular}
  \caption{The orientation of the three half wave plates $\theta_i$ (with $i=1,2,3$) and the  phase shift PS ($\phi$) for the 16 quantum states $\psi_{ij}$. The quantum success probabilities $P^Z_{exp}$ and $P^X_{exp}$ for measurements in the computational and Fourier basis respectively.}
\label{table2}
\end{table}

\textit{High-level QRACs $n=3$.---} Let us consider a second family of QRACs, on the form $3^{(d)}\rightarrow 1$ i.e. Alice encodes $x_0,x_1,x_2\in\{0,...,d-1\}$ into a quantum $d$-level system sent to Bob. Bob is supplied with a third measurement in addition to the bases $\{|l\rangle\}_l$ and $\{|e_l\rangle\}_l$. This measurement is choosen as a third MUB, $
|f_l\rangle=\frac{1}{\sqrt{d}}\sum_{k=0}^{d-1}\omega^{kl+\frac{k^2}{2}(1+\delta_d)}|k\rangle$ where $\delta_d=1$ if $d$ is an odd prime, and $\delta_d=0$ otherwise. 

Alice has $d^3$ different encodings but we begin by considering the $d$ encoding states on the form $|\psi_{00a}\rangle$, where we make the ansatz
\begin{equation}
|\psi_{00a}\rangle=\frac{1}{N_{3,d}}\left(|0\rangle+(r+it)|e_0\rangle+(r-it)|f_a\rangle\right)\\\label{3to1encoding}
\end{equation}
where $r$ and $t$ are real parameters and $N_{3,d}=\left(1+\frac{4r}{\sqrt{d}}+2r^2+2t^2+2\Re\left(\xi_a\right)\left(r^2-t^2\right)+4rt\Im\left(\xi_a\right)\right)^{1/2}$ is the normalization where we have introduced $\xi_a=\frac{1}{d}\sum_{k=0}^{d-1}\omega^{ak+\frac{k^2}{2}(1+\delta_d)}$.
This ansatz is made for symmetry reasons soon to appear.

We now claim that given an arbitrary encoding state on the form $|\psi_{00a}\rangle$, if Bob performs a measurement either in the basis $\{|e_l\rangle\}_l$ or in the basis $\{|f_l\rangle\}_l$, the probability of a obtaining the correct value of the corresponding $d$-levels $x_1$ and $x_2$ are always the same. To show this, we only need to expand the probability expressions $P_1(x_1=0)$ and $P_2(x_2=a)$ given below
\begin{eqnarray}\label{P00}
|\langle e_0|\psi_{00a}\rangle|^2
=\frac{1}{N_{3,d}^2}\left|\frac{1}{\sqrt{d}}+r+it+(r-it)\xi_a\right|^2\\\label{P1a}
|\langle f_a|\psi_{00a}\rangle|^2
=\frac{1}{N_{3,d}^2}\left|\frac{1}{\sqrt{d}}+(r+it)\overline{\xi}_a+r-it\right|^2
\end{eqnarray}
where $\overline{\xi}_a$ denotes the conjugate of $\xi_a$. Observing that the real parts inside the modulus squared in both \eqref{P00} and \eqref{P1a} are the same, and that the imaginary parts differ by a sign, it follows that the success probabilities of Bob are the same, $P_{1}(x_1=0)=P_{2}(x_2=a)$. This symmetry property motivates our ansatz in \eqref{3to1encoding}.

We now require that for given encoding state, Bob successfully recovers any $x_j$ with the same success probability. To enforce this, we compute the success probability $P_0(x_0=0)$ of Bob.
\begin{equation}\label{Pcomp0}
P_0(x_0=0)\equiv|\langle 0|\psi_{00a}\rangle|^2=\frac{1}{N_{3,d}^2}\left(1+\frac{2r}{\sqrt{d}}\right)^2
\end{equation}
Since we have shown that $P_1(x_1=0)=P_2(x_2=a)$ for all $a,d,r,t$, we need only to require that $P_0(x_0=0)=P_1(x_1=0)$. Realizing this constraint using \eqref{Pcomp0} and \eqref{P00}, one will find a set of values $(r,t(r))$ as function of $d$ and $a$ such that the state $|\psi_{00a}\rangle$ necessarily fulfills that the Bob's success probability is the same independent of his measurement. One is lead to a qudratic equation in $t$, $At^2+Bt+C=0$ with coefficients
\begin{gather}
A=1+|\xi_a|^2-2\Re\left(\xi_a\right)\\
B=2\Im\left(\xi_a\right)\left(2r+\frac{1}{\sqrt{d}}\right)\\
C=
r^2\left(|\xi_a|^2+2\Re\left(\xi_a\right)+1-\frac{4}{d}\right)+\frac{2r}{\sqrt{d}}\left(\Re\left(\xi_a\right)-1\right)+\frac{1}{d}-1
\end{gather}
with the familiar solution, now parametrized by $r$
\begin{equation}\label{opt}
t_{\pm}(r)=\frac{-B\pm\sqrt{B^2-4AC}}{2A}
\end{equation}
For given $d$ and $a$, and arbitrary choice of $r$ (as long as the corresponding $t_{\pm}$ is real), there is a corresponding QRAC with success probability independent of Bob's choice of measurement. However, we are interested in QRACs with high success probabilities and it is therefore natural to for given $d$ and $a$, choose the value of $r$ that optimizes Bob's success probability. A key observation is that the success probability depends on the value of the encoding $a$. In contrast to the case of $2^{(d)}\rightarrow 1$ QRACs, the average success probability and the worst case success probability are not the same for this family of $3^{(d)}\rightarrow 1$ QRACs.

Before computing the average success probability we are required to determine the arbitrary encoding states on the form $|\psi_{x_0x_1x_2}\rangle$ which we so far have not considered. However, general encoding states with good success probabilities can be generated from the encoding states $|\psi_{00a}\rangle$ by acting with the $X$ and $Z$ operators. We choose to define the action of these operators on the encoding $x_0x_1x_2=00a$ as
\begin{gather}
X^{\alpha}Z^{\beta}: x_0=0\rightarrow x_0=\alpha\\
X^{\alpha}Z^{\beta}: x_1=0\rightarrow x_1=\beta\\
 X^{\alpha}Z^{\beta}: x_2=a\rightarrow x_2=\beta+a-(1+\delta_d)\alpha
\end{gather}
where the sums are taken modulo $d$. It can be proven that if Bob measures on the encoding state
\begin{equation}
|\psi_{x_0x_1 f(x_0,x_1,d)}\rangle = X^{x_0}Z^{x_1}|\psi_{00x_2}\rangle
\end{equation}
where $f(x_0,x_1,d)=x_1+x_2-(1+\delta_d)x_0$, then the success probability is the same as if he measured on $|\psi_{00x_2}\rangle$ i.e. the success probability is invariant under actions on the encoding state $|\psi_{00a}\rangle$ with $X^\alpha Z^{\beta}$ for $\alpha,\beta\in\{0,...,d-1\}$. Evidently, using actions of $X$ and $Z$, it is impossible to map any $|\psi_{00x_2}\rangle$ to $|\psi_{00x_2'}\rangle$ for $x_2\neq x_2'$ and thus we are guaranteed to generate $d^3$ distinct encoding states with the above method. We can therefore compute the average quantum success probability, which by the above argument together with the probability expression \eqref{Pcomp0} can be written as
\begin{equation}\label{QRAC3}
p_{3,d}^Q=\frac{1}{d}\sum_{k=0}^{d-1} \max_{r,t_{\pm}}\frac{1}{N_{3,d}(k,r,t_{\pm})^2}\left(1+\frac{2r}{\sqrt{d}}\right)^2.
\end{equation}

In figure \ref{fig:1} we plot the ratio of the success probability of the $3^{(d)}\rightarrow 1$ QRACs and the RACs from equation \eqref{C2}. The quantum advantage is maximal for $d=13$ at which $p_{3,d}^Q/p_{3,d}^C\approx 1.224$ which is signficantly larger than the quantum advantage of the known two-level QRAC $3^{(2)}\rightarrow 1$ for which $p_{3,2}^Q/p_{3,2}^C\approx 1.052$. Evidently, the quantum advantage can be more than four-doubled by considering a high-level QRAC.

\textit{Discussion of optimality. ---} An important question is whether the average success probabilities, $p^Q_{n,d}$, that we have found are optimal, or if there exists other states and measurement which lead to better results. To investigate this, we have performed optimization using semidefinite programs (SDPs) \cite{VB96}, alternating between optimizing over states and measurements, to find lower bounds on $p^Q_{n,d}$. We have investigated cases $n=2$ ($n=3$) for $d=2,...,15$ ($d=2,...,8$) for all of which the obtained lower bound, up to numerical precision, coincides with our presented results. To find upper bounds on $p^Q_{n,d}$, we have used the intermediate, almost quantum, level of the Navascues-Vertesi hierarchy of dimensionally constraint quantum correlations \cite{NV14}. Due to computational limitations, we have only studied the particular cases with $n=2$ and $d=2,3,4,5$, for which we found that the obtained upper bounds coincide with the lower bounds, hence proving the optimality of these particular QRACs. We also investigated our $3^{(3)}\rightarrow 1$ QRAC by the same method. Here, the lower bound, $p^Q_{n,d}\approx 0.6971$, is somewhat smaller than the obtained upper bound, $0.6989$. Evidently, to determine the optimality of this particular QRAC, we need to run higher levels of the Navascues-Vertesi hierarchy, which is compuationally demanding. The optimality of our $3^{(3)}\rightarrow 1$ QRAC remains an open problem.

\textit{Conclusions.---} In this letter, we have investigated both classical and quantum random access codes with $d$-level communication. We derived the average success probability of what we strongly believe is the optimal RAC, along with two families of QRACs showing that they provide a significant gain in success probability over the RACs. Our results show that high-level QRACs enable significantly larger advantages over the corresponding RACs than what can be achieved with QRACs using two-level systems. The QRACs presented here can be applied to a variety of problems in quantum information. Nevertheless, this investigation also leaves many open questions on the properties of random access codes: (1) find an efficient way to generalize our QRACs to arbitrary $n$. (2) RACs and QRACs can be considered in other scenarios e.g. if Bob wants to access a larger subset of Alice's data and/or when Alice is allowed to communicate more than one $d$-level. (3) Consider scenarios with the dimension of Alice's data set not being equal to the dimension of the communicated system i.e. Alice holds a string of $n$ $d$-levels while communicating a single classical (quantum) $d'$-level to Bob with $d\neq d'$. (4) Our approach to QRACs is based on MUBs, however it is interesting to ask if this is always an optimal choice of basis for any $n^{(d)}\rightarrow 1$ QRAC, or if there are bases that result in QRACs with better performance.

\textbf{Acknowledgements} The authors thanks David Gaharia, Ingemar Bengtsson, Marcus Grassl and Marcin Pawlowski. This project was supported by the Swedish Research Council, ADOPT,  CAPES (Brazil),  and ERC-QOLAPS grants.

\end{document}